\documentclass[10pt,twocolumn,letterpaper]{article}

\usepackage[pagenumbers]{iccv} % To force page numbers, e.g. for an arXiv version
\usepackage[accsupp]{axessibility}  % Improves PDF readability for those with disabilities.
\usepackage[dvipsnames,table]{xcolor}

\definecolor{iccvblue}{rgb}{0.21,0.49,0.74}
\usepackage[pagebackref,breaklinks,colorlinks,allcolors=iccvblue]{hyperref}
\usepackage{multirow}
\usepackage{booktabs}   % in your preamble
\usepackage{tabularx}    % in your preamble
\usepackage{stfloats}

\title{Weakly Supervised Intracranial Aneurysm Detection and Segmentation in MR angiography via Multi-task UNet with Vesselness Prior}

\author{
    Erin Rainville$^{1}$\thanks{Correspondence to: \texttt{e\_ainvil@live.concordia.ca}.} \hfill
    Amirhossein Rasoulian $^{2}$ \hfill
    Hassan Rivaz$^{3}$ \hfill
    Yiming Xiao$^{1}$ \\
    $^1$Department of Computer Science and Software Engineering, Concordia University, Montréal, Canada \\
    $^2$NeuroRx Research, Montréal, Canada
    \\
    $^3$Department of Electrical and Computer Engineering, Concordia University, Montréal, Canada
}

\begin{document}
\maketitle
\begin{abstract}
Intracranial aneurysms (IAs) are abnormal dilations of cerebral blood vessels that, if ruptured, can lead to life-threatening consequences. However, their small size and soft contrast in radiological scans often make it difficult to perform accurate and efficient detection and morphological analyses, which are  critical in the clinical care of the disorder. Furthermore, the lack of large public datasets with voxel-wise expert annotations pose challenges for developing deep learning algorithms to address the issues. Therefore, we proposed a novel weakly supervised 3D multi-task UNet that integrates vesselness priors to jointly perform aneurysm detection and segmentation in time-of-flight MR angiography (TOF-MRA). Specifically, to robustly guide IA detection and segmentation, we employ the popular Frangi's vesselness filter to derive soft cerebrovascular priors for both network input and an attention block to conduct segmentation from the decoder and detection from an auxiliary branch. We train our model on the Lausanne dataset with coarse ground truth segmentation, and evaluate it on the test set with refined labels from the same database. To further assess our model's generalizability, we also validate it externally on the ADAM dataset. Our results demonstrate the superior performance of the proposed technique over the SOTA techniques for aneurysm segmentation (Dice = 0.614, 95\%HD =1.38mm) and detection (false positive rate = 1.47, sensitivity = 92.9\%). 

\end{abstract}    
\section{Introduction}
\label{sec:intro}
An intracranial aneurysm (IA) is a cerebrovascular disorder characterized by the abnormal, localized bulging of a cerebral artery due to a weakness in the vessel wall. It affects approximately 3\% of the global population and often remains undetected due to its asymptomatic nature~\cite{ztaufique,asliwczynski}. When an aneurysm ruptures, it is the leading cause of subarachnoid hemorrhage (SAH), a life-threatening type of stroke~\cite{ztaufique}. SAH has a 35$\sim$45\% mortality rate and nearly half of the survivors experience significant long-term neurological disabilities~\cite{asliwczynski,gboulouis}. Recent studies~\cite{ltvdkamp,njuchler,ywang} emphasize that the morphology of aneurysms, especially shape irregularities and growth detected during follow-ups, are critical in predicting rupture risk. Therefore, there is a clear need for early detection and precise segmentation of unruptured intracranial aneurysms (UIAs) to better manage preventative aneurysm treatment.

For the diagnosis and analysis of intracranial aneurysms, computed tomography angiography (CTA) and magnetic resonance angiography (MRA) are two primary imaging modalities commonly adopted ~\cite{amhsailer,zzhou}. While CTA is faster with high resolution for diagnostic accuracy ~\cite{mclemence}, it exposes patients to risks of ionizing radiation and potential adverse reactions to iodinated contrast agents~\cite{xchen}. On the other hand, time-of-flight (TOF) MRA  avoids exposure to radiation or iodinated contrast and is better suited for long-term monitoring and follow-up assessments of UIAs~\cite{xchen}. However, due to softer vascular contrast and lower spatial resolution, IA diagnostic accuracy may suffer~\cite{mclemence}. Traditionally, radiologists manually identify and measure UIAs by annotating large imaging volumes slice by slice. This is a tedious and time-consuming task and it has been estimated that approximately 10\% of all UIAs are missed during standard screening ~\cite{pmwhite}. To address these limitations, recent advancements in deep learning (DL) have enabled automated extraction and analysis of complex features from medical images, enhancing the efficiency and accuracy of UIA detection and segmentation tasks~\cite{dshen}. This is particularly beneficial for safer MRA-based UIA diagnosis and analysis. The current DL approaches for UIA assessment with TOF-MRA face two main challenges. First, the small size, sparse occurrence in a brain volume, and subtle morphological features can introduce strong issues of class-imbalance and feature localization. Second, there is a lack of large, well-annotated public MRA datasets for UIA segmentation due to the cost of expert manual labels, making it the bottleneck to develop and validate relevant DL methods.

To address the clinical need and challenges aforementioned, we introduce the Vessel-Prior UNet (VP UNet), a novel 3D multi-task UNet that integrates spatial vesselness priors into both UIA segmentation and detection based on weak segmentation ground truths during training. Our contribution is three-fold: \textbf{First}, as UIAs are a pathology of the blood vessels, we propose to leverage the popular and robust Frangi's vesselness filter~\cite{affrangi} to derive soft spatial priors of blood vessels to guide UIA feature learning. \textbf{Second}, we design a novel multi-task UNet for joint UIA detection and segmentation to benefit from their synergy. Specifically, we integrate the Frangi vesselness priors through shared feature learning and attention gating; while producing segmentation masks at the UNet decoder, UIA detection is obtained with the joint use of multi-scale information from both the bottleneck and the decoder. \textbf{Lastly}, we used coarse segmentation ground truths (i.e., simple spheres) and test time augmentation (TTA) to mitigate the burdens in refined manual labeling, and we validated our proposed method against the state-of-the-art (SOTA) methods on two different datasets.

\section{Related Work}
\label{sec:relatedwork}

Deep learning methods have become the standard in medical image analysis. Two recent large-scale reviews~\cite{zzhou,zwen} show that CNNs remain the most commonly used model family for intracranial aneurysm detection, and that UNet variants continue to serve as the backbone for most segmentation pipelines. This trend is also reflected in the Aneurysm Detection And segMentation (ADAM) Challenge, organized alongside MICCAI 2020, where 72\% of participating methods used UNet variants, including the top-performing submissions for both detection and segmentation tasks~\cite{kmtimmins}. We adopt a UNet-based framework in our study to remain aligned with these proven architectural choices. We also used the ADAM dataset to externally validate the generalizability of our weakly supervised model.

Among more recent approaches for UIA detection and segmentation, Ham et al.~\cite{sham} proposed a novel skeleton-based network trained on their in-house TOF-MRA data, where vessel segmentations are first computed to extract the vasculature from MRA scans. The vessel-aligned 3D patches are then sampled along the vessel skeleton and passed through a 3D UNet with an auxiliary classifier. Their approach leverages deterministic vessel segmentation as a hard spatial constraint, enforcing that both training and inference occur only along segmented vessels, which helps address class imbalance and focus learning on vascular regions. While effective, this hard constraint depends heavily on accurate vessel segmentation; if parts of the vasculature are mis-segmented, aneurysms located outside the segmented skeleton cannot be detected~\cite{aceballos}. In contrast, our proposed method incorporates a soft vesselness prior from the Frangi vesselness filter~\cite{affrangi}, that is passed as an additional input to the network and to an attention gate. This allows the network to learn how vessel information should influence classification and segmentation jointly, even with imperfect vessel enhancement.

Deep learning methods for medical segmentation often face the challenge of limited well-annotated data, since obtaining precise voxel-wise ground truths, particularly for small intracranial aneurysms, is labour-intensive and demands significant clinical expertise. To address this, weakly supervised segmentation has emerged as a practical alternative, relying on coarser annotations (e.g., bounding boxes, scribbles, and rough contours) that are faster and less costly to produce~\cite{arasoulian,mrajchl,gyang}. For instance, Di Noto et al.~\cite{tdinoto} proposed the use of spherical annotations that fully enclose aneurysms as a form of weak labels for UIA segmentation. While less precise than full segmentation, these annotations can be created four times more efficiently. Building on this, we adopt coarse spherical ground truths of UIAs from TOF-MRA scans to train our model.

By integrating weak labels to encourage scalable generation of datasets, a multi-task framework to enhance feature generalization, and soft vesselness prior to improve robustness against imperfect segmentation, our proposed VP UNet uniquely addresses key limitations highlighted in previous works.
\section{Data Processing}
\label{sec:materials}
In this study, we used two publicly available TOF-MRA datasets of UIA segmentation. First, Di Noto et al.'s~\cite{tdinoto} Lausanne dataset contains 284 subjects (157 patients with one or more aneurysms, and 127 healthy controls). Among these, 246 subjects have weak labels (simplistic spheres) that completely enclose aneurysms while 38 subjects have precise voxel-wise segmentations. Second, for external validation, Timmins et al.'s~\cite{kmtimmins} ADAM dataset consists of 113 subjects, with 93 aneurysm-positive patients with voxel-wise UIA segmentations and 20 healthy controls. Note that both datasets only captured mid-slabs of the brain with the main brain vasculatures, while they imaged the brain in different head orientations, resulting in larger field of view for the Lausanne dataset. 

To prevent data leakage, all dataset splits occurred at the subject level. The Lausanne dataset's 38 voxel-wise segmented cases served as our internal test set, while the remaining 246 weakly-labelled cases were randomly split into our training set (90\%, 222 subjects) and validation set (10\%, 24 subjects). The full ADAM cohort was reserved as an external test set to evaluate model generalization.

\subsection{Image Pre-processing}
The Lausanne dataset underwent four preprocessing steps. First, skull stripping was performed using the FSL Brain Extraction Tool (BET)~\cite{smsmith}. Second, we applied N4 bias field correction with SimpleITK~\cite{njtustison}. Third, all images were resampled to a uniform median resolution of $[0.39, 0.39, 0.55]$ mm$^3$. Lastly, a probabilistic vessel atlas developed from multi-center MRA datasets~\cite{pmouches} was registered to each subject’s structural T1w MRI, and subsequently to the TOF-MRA volume, using ANTS~\cite{bbavants}. This process enabled anatomical landmark mapping critical for anatomical patch extraction. For consistency, the ADAM dataset was preprocessed in the same manner.

\subsection{Patch Extraction and Vesselness Maps}
Training samples from Lausanne’s weak labels were prepared following Di Noto et al.'s publicly available patch extraction pipeline~\cite{tdinoto}. $64\times64\times64$ voxel patches were extracted for efficient computation and then processed with z-normalization. We extracted approximately 50 negative patches (no aneurysms) per subject using a balanced selection of vessel-like, landmark-centered, and random patches. Then, eight positive patches with different offsets were extracted for each aneurysm. To mitigate class imbalance, positive patches underwent extensive data augmentation, including intensity-based transformations (Gaussian noise injection, contrast adjustments, and intensity shifts) and geometric augmentations (rotations, flips, and zooming). Two to five augmentations were randomly applied to each patch. During model training, a weighted random sampler increased the likelihood of selecting positive patches.

During inference, we followed  Di Noto et al.'s ``anatomically informed" patch extraction method, which extracts inference patches around precise locations of the vasculature that have a high probability of aneurysms using 20 landmarks defined on the aligned probabilistic vessel atlas~\cite{tdinoto,pmouches}. Approximately 50 inference patches were extracted per subject and they underwent the same pre-processing steps as the training set.

\begin{figure}
    \centering
    \includegraphics[width=\linewidth]{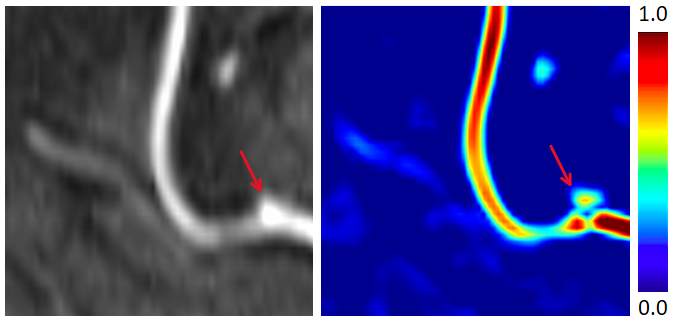}
    \caption{Image patch with aneurysm (left) and corresponding vesselness map in jet colormap (right). The location of the aneurysm is indicated with red arrows.} \label{fig1}
\end{figure}

Each image patch was complemented by a corresponding vesselness map (Fig.~\ref{fig1}). The original image patch was filtered with a Hessian matrix, then the Frangi vesselness function was applied to its eigenvalues to detect tubular and blob-like structures within the image~\cite{affrangi}. Here, we used the default parameters of $\sigma$ = 1.0, $\alpha_1$ =0.5, $\alpha_2$ =2.0 for the Frangi vesselness function, which we found to offer the best results based on our empirical observation.

\subsection{Segmentation Post-Processing}
During inference, the models predict the voxel-wise segmentation of the anatomically-informed test patches~\cite{tdinoto}. Three post-processing steps were employed to enhance the results, particularly as a result of using weak segmentation labels. \textbf{First}, we applied test-time augmentations (simple geometric transformations of flipping and rotating 90 degrees) as it has been shown to produce more robust prediction results~\cite{ikim}, and could mitigate the impact of inconsistency in weakly annotated training samples. The average of those results is used as the prediction region. \textbf{Second}, as segmentation results could include some sporadic labels as false positives due to factors like image noise, we thus remove any connected regions whose size is below 5 voxels, with the assumption that an aneurysm should be larger than this volume. \textbf{Third}, we fill any holes within the connected region to produce a final predicted aneurysm segmentation.

\section{Network Architecture}
\label{sec:network}
\begin{figure*}
    \centering
    \includegraphics[width=.98\textwidth]{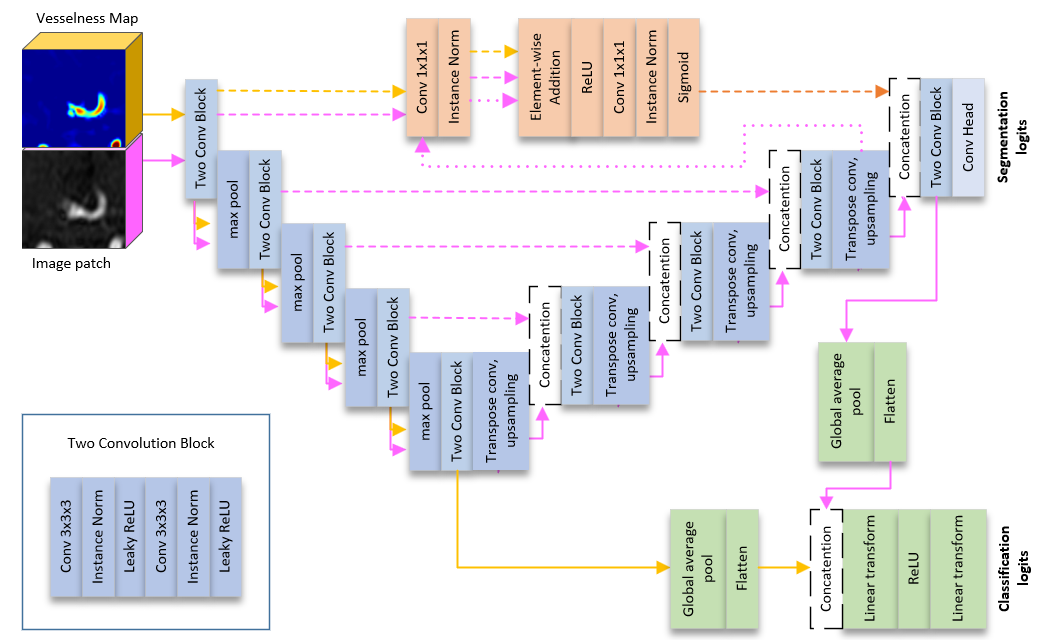}
    \caption{Network architecture of the proposed multi-task VP UNet. It is composed of a 3D UNet (blue), an Attention Block in the top skip connection (orange), and the MT-UNet-based auxiliary classification branch (green). The lines depict the passing of information: the image patch (pink), the vesselness map (yellow), and the skip connections (dashed lines). The logits are used for the loss function. } 
    \label{fig2}
\end{figure*}
Our model was partially inspired by the multi-task (MT) UNet framework from ~\cite{hzhu} and attention UNet~\cite{ooktay}. The full architecture is depicted in Fig.~\ref{fig2}. It jointly processes a $64\times64\times64$ voxel image patch and its corresponding vesselness map in a single 3D UNet encoder, then branches into classification and segmentation decoders. This design maximizes parameter sharing and computational efficiency, while guiding the network toward vascular structures.

\subsection{Shared Encoder}
We adopt the four-layer 3D UNet~\cite{ocicek} encoder, which applies a two-convolution block (two  successive $3\times3\times3$ convolutions each followed by Instance Norm and LeakyReLU) then a $2\times2\times2$ max pool with stride of 2 at each layer. Crucially, the image patch and its vesselness map are not concatenated, instead they both traverse the same encoder path, keeping parameter count nearly unchanged, and ensuring that vesselness priors are inherently embedded from the earliest layers. At the bottleneck, a final two-convolution block fuses high-level features before splitting into two task-specific branches. This shared encoder for both the MRA image and the vesselness map ensures that both classification and segmentation tasks benefit from the same multi-scale representations, with a focus on encoding vessel-related image features.

\subsection{Classification Head}
Drawing from the auxiliary classification head of the MT‑UNet framework ~\cite{hzhu}, the classification branch aggregates the global average-pooled vesselness features from the bottleneck and the final up-sampled image patch from the UNet decoder, then concatenates them into a single vector. A dropout layer (droput rate = 20\%) precedes a fully connected layer with ReLU activation, followed by a final linear layer that outputs patch-wise aneurysm detection logits. This fusion of encoder and decoder information at different scales preserves both the context and high-resolution detail.

\subsection{Segmentation Decoder and Attention Gating}
The segmentation branch follows a standard 3D UNet decoder with four up-sampling stages, each using a $2\times2\times2$ transposed convolution with a stride of 2 to restore spatial resolution. At each stage, corresponding encoder features are concatenated via skip connections. To enhance the focus of aneurysm segmentation close to regions near the blood vessels, at the final layer of the decoder, we insert an Attention Block~\cite{ooktay} rather than a simple skip connection. It takes as input the encoder image features, the encoder vesselness map features, and the decoder's gating signal. Each input is mapped by a $1\times1\times1$ convolution and Instance Norm, then they are summed together and run through another $1\times1\times1$ convolution, Instance Norm, and sigmoid layer to yield an attention map. The map modulates the encoder features before concatenation, applying soft attention on vessel-rich regions~\cite{ooktay,wehab}. This output is passed through the segmentation head, a $3\times3\times3$ convolution, to produce the voxel-wise aneurysm logits.

\subsection{Loss Function}

We train our multi-task network with a joint classification-and-segmentation objective: 
\begin{equation}
    \label{eq:total_loss}
    \mathcal{L} = \phi L_{F} + (1-\phi)(\beta L_{GD} + (1 - \beta) L_{CE})
\end{equation}
where $\phi\in[0,1]$ is the trade-off between the two tasks, we fixed $\phi=0.3$ empirically. The classification term is an $\alpha$-balanced focal loss~\cite{tylin}: 
\begin{equation}
    \label{eq:focal_loss}
    L_F = -\alpha(1-p_C)^{\gamma}log(p_C)
\end{equation}
where $p_C$ is the predicted probability of the patch-wise aneurysm class. $\alpha=0.25$ and $\gamma=2.0$ are the default hyperparameters of the sigmoid focal loss. 

The segmentation loss itself is a mixture of generalized Dice~\cite{chsudre} and cross-entropy:
\begin{equation}
    \label{eq:gd_loss}
    L_{GD} = 1 - 2 \times \sum_c{\omega_C\frac{p_S \odot g_S}{p_S + g_S}}
\end{equation}
\begin{equation}
    \label{eq:ce_loss}
    L_{CE} = -\sum_c{g_S \cdot log(p_S)}
\end{equation}
where $g_S$ and $p_S$ are ground-truth and predicted probabilities of a pixel. $\omega_C$ is set inversely proportional to the class's frequency and $\beta\in[0,1]$ balances the two segmentation terms. We chose $\beta=0.5$ empirically. This combined objective encourages both accurate aneurysm detection (through the focal term) and precise mask overlap (through Dice and CE), improving performance under severe class imbalance.

 \begin{table*}
  \centering
  \begin{tabularx}{.9\textwidth}{@{}l *{4}{>{\raggedright\arraybackslash}X}@{}}
    \toprule
    \multirow[b]{2}{*}{\textbf{Model}}
      & \multicolumn{2}{c}{\textbf{Internal Test Set (Lausanne)}}
      & \multicolumn{2}{c}{\textbf{External Test Set (ADAM)}} \\
    \cmidrule(lr){2-3} \cmidrule(lr){4-5}
      & \textbf{FP rate \textcolor{red}{$\downarrow$}}
      & \textbf{Sensitivity \textcolor{green!70!black}{$\uparrow$}}
      & \textbf{FP rate \textcolor{red}{$\downarrow$}}
      & \textbf{Sensitivity \textcolor{green!70!black}{$\uparrow$}} \\
    \midrule
    3D U-Net                   & 2.778±1.565 & 0.933±0.165 & 2.012±1.460 & \textbf{0.854±0.329} \\
    MT-UNet        & 1.944±1.201 & 0.786±0.383 & \underline{1.310±1.422} & 0.533±0.462 \\
    Swin UNETR          & \underline{1.750±1.277} & \textbf{0.971±0.116} & 1.429±1.383 & 0.777±0.366 \\
    ResUnet             & 2.444±1.536 & \underline{0.962±0.127} & 1.607±1.195 & 0.822±0.349 \\
    \rowcolor{gray!15} 
    VP UNet (Ours)            & \textbf{1.472±1.093} & 0.929±0.212 &  \textbf{1.143±1.216} &  \underline{0.828±0.337 }\\
    \bottomrule
  \end{tabularx}
  \caption{Comparing baselines and our proposed model (in grey) for detection performance on internal and external test sets (mean±std). Best results in bold, second best results underlined. All models have TTA post-processing.}
  \label{tab:detect_results}
\end{table*}

\begin{table*}
  \centering
  \begin{tabularx}{.99\textwidth}{@{}l *{3}{>{\centering\arraybackslash}X} *{3}{>{\centering\arraybackslash}X}@{}}
    \toprule
    \multirow[b]{2}{*}{\textbf{Model}}
      & \multicolumn{3}{c}{\textbf{Internal Test Set (Lausanne)}}
      & \multicolumn{3}{c}{\textbf{External Test Set (ADAM)}} \\
    \cmidrule(lr){2-4}\cmidrule(lr){5-7}
      & \textbf{DICE \textcolor{green!70!black}{$\uparrow$}}
      & \textbf{IoU \textcolor{green!70!black}{$\uparrow$}}
      & \textbf{95-HD \textcolor{red}{$\downarrow$}}
      & \textbf{DICE \textcolor{green!70!black}{$\uparrow$}}
      & \textbf{IoU \textcolor{green!70!black}{$\uparrow$}}
      & \textbf{95-HD \textcolor{red}{$\downarrow$}} \\
    \midrule
    3D U-Net                      & \underline{0.587±0.105} & 0.425±0.102 & \textbf{1.336±0.532}
                              & 0.461±0.190 & 0.321±0.160 & 1.660±0.831 \\
    MT-Unet                   & 0.514±0.190 & 0.367±0.153 & 1.852±1.225 
                              & 0.408±0.235 & 0.284±0.192 & 2.114±1.226 \\
    Swin UNETR                &\underline{0.587±0.153} & \underline{0.432±0.135} & 1.492±0.726
                              & \textbf{0.503±0.184} & \textbf{0.357±0.164} & \textbf{1.584±0.693} \\
    ResUnet                   & 0.571±0.150 & 0.418±0.131 & 1.496±1.043 
                              & 0.470±0.200 & 0.332±0.176 & \underline{1.619±0.864} \\
    \rowcolor{gray!15} 
    VP UNet (Ours)              & \textbf{0.614±0.137} &  \textbf{0.456±0.128} &  \underline{1.379±0.867} 
                              & \underline{0.489±0.203} & \underline{0.349±0.177 }& 1.635±0.908 \\
    \bottomrule
  \end{tabularx}
  \caption{Comparing baselines and our proposed model (in grey) for segmentation performance on internal and external test sets (mean\,±\,std). 95-Hausdorff is in mm. Best results in bold, second best results underlined. All models have TTA post-processing.}
  \label{tab:seg_results}
\end{table*}

\section{Evaluation Metrics}
The proposed model and the comparison baselines were evaluated with detection metrics (Table~\ref{tab:detect_results}), including false positive (FP) rate and sensitivity, and segmentation metrics (Table~\ref{tab:seg_results}) of Dice coefficient, Intersection over Union (IoU) and 95\% Hausdorff Distance (95-HD). A successful detection is defined as any intersection between a predicted region and the true segmentation region. While the proposed network processes image patches, within each subject, we calculated the metrics per aneurysm and then averaged the results to obtain per-subject metrics. Note that segmentation metrics were only calculated for true positive aneurysm detections. For the external validation dataset (i.e., ADAM), no samples from it were used in model training to ensure the proper assessment on model generalizability to different scanners and imaging protocols.

\section{Experimental Setup and Results}
\label{sec:results}

\subsection{UIA Detection and Segmentation}
\begin{figure*}
  \centering
  \includegraphics[width=\linewidth]{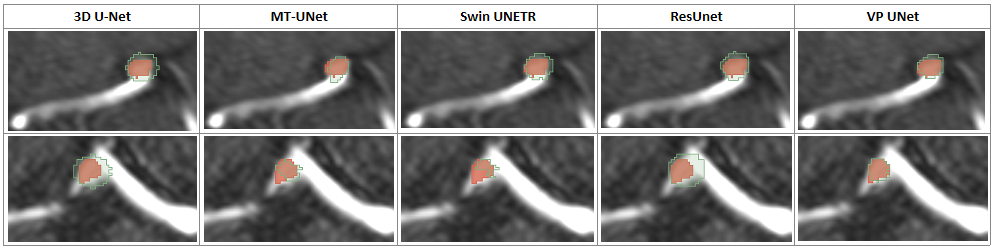}
   \caption{Qualitative comparison of segmentation results for two different patients (one patient per row) from the Lausanne dataset. The \textcolor{red}{red} label is the ground truth, and the \textcolor{green!60!black}{green} label represents the automatic segmentation.}
   \label{fig:qualitative_segs}
\end{figure*}

We evaluated the performance of our proposed vessel-guided multi-task UNet against several established baselines: the 3D U-Net~\cite{ocicek}, 3D adaptation of the multi-task UNet~\cite{hzhu}, Swin UNETR~\cite{ahatamizadeh}, and 3D ResUNet~\cite{zzhang}. All models were trained and evaluated using the same dataset splits, pre-processing pipeline, and segmentation post-processing (including TTA) to ensure a fair comparison. The data was split at the subject level to avoid data leakage, with a balanced distribution of positive and negative patches in both training and validation sets. Models were trained with a batch size of 24 using the AdamW optimizer (initial learning rate 0.001, decayed by 20\% every 5 steps). Training continued for up to 100 epochs, with early stopping triggered if the validation loss plateaued (change less than 0.001) over 10 consecutive epochs.

The UIA Detection performance is summarized in Table~\ref{tab:detect_results}, separated by dataset to assess generalization. The best and second-best results for each metric are highlighted in bold and underlined, respectively. Our proposed model achieved the best false positive rate on both internal and external test sets, reducing the metric by 0.47 (internal dataset) and 0.17 (external dataset) compared to the next best models. This suggests that our model is more robust at discriminating true aneurysms from false positives, likely due to the integration of a soft vesselness prior that guides the network toward plausible vascular regions. In terms of sensitivity, our model performed comparably to other methods, with most models missing only 2–4 aneurysms during inference ($\sim$10 aneurysms externally).
However, because the test sets contain only a small number of aneurysms, even when models miss a similar number of lesions, each false negative produces a substantial change in the reported sensitivity.

On the other hand, UIA segmentation performance is reported in Table~\ref{tab:seg_results}. Our model achieved the best Dice and IoU scores on the internal test set, with improvements of 0.03 Dice and 0.02 IoU over the next best model. On the external test set, it ranked second, trailing the top model by just 0.01 for both Dice and IoU. All models exhibited reduced segmentation accuracy on the external dataset, highlighting the challenges in generalizing across datasets, particularly in data acquisition protocols. Notably, our model lagged in 95\textsuperscript{th}-percentile Hausdorff distance (95HD), with increases of 0.04 mm (internal) and 0.05 mm (external) compared to the best-performing models, suggesting room for improvement in boundary precision. 

Qualitative results for two patient cases are shown in Fig.~\ref{fig:qualitative_segs}, illustrating the segmentation quality and differences across DL models. 

\subsection{Ablation Studies}

In addition to benchmarking against existing architectures, we conducted ablation studies to investigate the contribution of specific components in our proposed model, particularly the architectural integration of vesselness priors at the UNet encoder and the attention block, as well as the use of test-time augmentation (TTA). Each variant was evaluated for both aneurysm detection and segmentation tasks to provide detailed insights into their impacts. To help guide the readers, Table~\ref{tab:ablation-vesselness} summarizes the architectural differences between the ablated model variants.

For aneurysm detection, Table~\ref{tab:detect_ablation} illustrates the inherent trade-off between false positive rate and sensitivity. Models that aggressively reduce FP rate, such as the variant with vesselness guidance limited to the encoder, tend to sacrifice sensitivity. Conversely, models that maintain high sensitivity often exhibit elevated FP rates. Our final model (VP UNet) achieves the best overall balance, with the lowest FP rate (1.472) on the internal test set and second-lowest FP rate (1.143) on the external test set, while maintaining strong sensitivity (0.929 internal / 0.828 external). This suggests that the joint supervision, soft vesselness guidance, and TTA collectively help reduce misclassifications without under-detecting true aneurysms.

\begin{table}[bp]
\centering
    \begin{tabular}{lccc}
    \toprule
    \textbf{Model} & \textbf{Encoder} & \textbf{Attention Block} & \textbf{TTA}\\
    \midrule
    \textbf{Vessel Encoder}  & $\checkmark$  & $\times$ & $\checkmark$ \\
    \textbf{Vessel AttBlock} &  $\times$ & $\checkmark$  & $\checkmark$  \\
    \textbf{No TTA}          & $\checkmark$ & $\checkmark$  &  $\times$ \\
    \rowcolor{gray!15}
    \textbf{VP UNet (Ours)}    & $\checkmark$ & $\checkmark$  & $\checkmark$\\
    \bottomrule
    \end{tabular}
\caption{Ablation study on where the vesselness map is used as input and the impact of TTA post-processing. $\checkmark$ indicates the component is used in that model.}
\label{tab:ablation-vesselness}
\end{table}

For aneurysm segmentation, Table~\ref{tab:seg_ablation} shows that our final proposed model achieves the best Dice score (0.614) and IoU (0.456) on the internal test set, and on the external test set (Dice = 0.489, IoU = 0.349), indicating a strong segmentation capability. While its 95\textsuperscript{th} percentile Hausdorff distance (95-HD) is slightly higher than the best performing models, the margin is small and again highlights an opportunity for further optimization in boundary refinement.

Comparing our final model to the variant without TTA confirms the benefit of test-time augmentation: segmentation performance improved across all metrics, and the FP rate was also reduced. This supports our hypothesis that TTA helps compensate for the limited precision of weak spherical annotations by enhancing spatial consistency during inference.

Overall, these ablation results confirm our architectural choices. Fusing the vesselness prior at multiple levels and applying TTA at inference both contribute to more accurate and robust detection and segmentation of aneurysms in weakly annotated TOF-MRA data.

 \begin{table*}[bp]
  \centering
  \begin{tabularx}{.9\textwidth}{@{}l *{4}{>{\raggedright\arraybackslash}X}@{}}
    \toprule
    \multirow[b]{2}{*}{\textbf{Model}}
      & \multicolumn{2}{c}{\textbf{Internal Test Set (Lausanne)}}
      & \multicolumn{2}{c}{\textbf{External Test Set (ADAM)}} \\
    \cmidrule(lr){2-3} \cmidrule(lr){4-5}
      & \textbf{FP rate \textcolor{red}{$\downarrow$}}
      & \textbf{Sensitivity \textcolor{green!70!black}{$\uparrow$}}
      & \textbf{FP rate \textcolor{red}{$\downarrow$}}
      & \textbf{Sensitivity \textcolor{green!70!black}{$\uparrow$}} \\
    \midrule
    Vessel Encoder   & 2.361±1.512 & \textbf{0.948±0.148} & 1.583±1.246 & \textbf{0.848±0.322}\\
    Vessel AttBlock  & 1.611±1.208 & \underline{0.948±0.190} & \textbf{0.905±0.934} & 0.810±0.360\\
    No TTA           & \underline{1.500±1.143} & 0.943±0.159 & 1.274±1.158 & \underline{0.836±0.335}\\
    \rowcolor{gray!15} 
    VP UNet (Ours) & \textbf{1.472±1.093} & 0.929±0.212 & \underline{1.143±1.216} &  0.828±0.337\\
    \bottomrule
  \end{tabularx}
  \caption{Comparing different architecture compositions and our proposed model (in grey) for detection performance on internal and external test sets (mean±std). Best results in bold, second best results underlined.}
  \label{tab:detect_ablation}
\end{table*}

\begin{table*}[bp]
  \centering
  \begin{tabularx}{.99\textwidth}{@{}l *{3}{>{\centering\arraybackslash}X} *{3}{>{\centering\arraybackslash}X}@{}}
    \toprule
    \multirow[b]{2}{*}{\textbf{Model}}
      & \multicolumn{3}{c}{\textbf{Internal Test Set (Lausanne)}}
      & \multicolumn{3}{c}{\textbf{External Test Set (ADAM)}} \\
    \cmidrule(lr){2-4}\cmidrule(lr){5-7}
      & \textbf{DICE \textcolor{green!70!black}{$\uparrow$}}
      & \textbf{IoU \textcolor{green!70!black}{$\uparrow$}}
      & \textbf{95-HD \textcolor{red}{$\downarrow$}}
      & \textbf{DICE \textcolor{green!70!black}{$\uparrow$}}
      & \textbf{IoU \textcolor{green!70!black}{$\uparrow$}}
      & \textbf{95-HD \textcolor{red}{$\downarrow$}} \\
    \midrule
    Vessel Encoder      & 0.587±0.147 & 0.432±0.130 & \textbf{1.330±0.545} 
                        & 0.480±0.194 & 0.340±0.169 & \textbf{1.586±0.848} \\
                        
    Vessel AttBlock     & 0.563±0.124 & 0.406±0.108 & 1.421±0.650 
                        & \underline{0.472±0.195} & \underline{0.332±0.168} & \underline{1.617±0.837} \\
                        
    No TTA              & 0.567±0.191 & 0.418±0.159 & 1.467±0.919 
                        & 0.466±0.207 & 0.330±0.178 &  1.733±1.043 \\
    \rowcolor{gray!15} 
    VP UNet (Ours)        & \textbf{0.614±0.137} & \textbf{0.456±0.128} & \underline{1.379±0.867}
                        & \textbf{0.489±0.203} & \textbf{0.349±0.177 }& 1.635±0.908 \\
    \bottomrule
  \end{tabularx}
  \caption{Comparing different architecture compositions and our proposed model (in grey) for segmentation performance on internal and external test sets (mean\,±\,std). 95-Hausdorff is in mm. Best results in bold, second best results underlined.}
  \label{tab:seg_ablation}
\end{table*}

\section{Discussion}
\label{sec:discussion}

Intracranial aneurysm detection and segmentation remain challenging tasks because aneurysms constitute small structures and are sparsely distributed within 3D brain scans while closely resembling adjacent vascular structures. To date, few previous studies have specifically explored weakly supervised models for UIA segmentation.  Since no public finely annotated datasets exist beyond ADAM (which we used solely as an external test set), the ADAM Challenge results serve as a de facto upper bound for fully supervised performance. Notably, the best segmentation results from the ADAM Challenge~\cite{kmtimmins} reported a 0.64 Dice score and a 2.62mm 95-HD score with refined training labels. In comparison, our proposed VP UNet, trained only on coarse spherical labels plus vesselness priors, achieved a comparable 0.61 Dice score (0.49 externally) and a reduced 95-HD score of 1.38mm (1.64mm externally) demonstrating that weak supervision can approach fully supervised accuracy while dramatically reducing the manual labour of finely annotated datasets.  For aneurysm detection, our model demonstrated a sensitivity of 92.9\% on the internal dataset and 82.8\% on the external dataset, both of which surpass the ADAM Challenge's best reported sensitivity of 67\%. These findings are important given that both training and inference were conducted at the patch level rather than the subject level, allowing for fast inference results without compromising aneurysm detection. 

Our model builds upon the established UNet architecture, maintaining consistency with prior literature, while uniquely integrating soft anatomical priors in the form of vesselness maps derived from raw MRA images. Integrating the Frangi vesselness map provides important context by guiding the network’s attention to vessel-like structures where aneurysms occur. This substantially reduced false positives in non-vascular regions without sacrificing sensitivity. Unlike hard constraints (e.g., skeleton-based sampling, which can miss aneurysms if the vessel mask is incomplete), the soft vesselness prior allows the model to learn when to trust the vessel features, making it robust even if the vessel filter is imperfect. Another factor in our model’s superior performance is the use of TTA during inference. By averaging predictions over multiple orientations of the input, we obtained more robust and stable segmentation results. TTA is a well-known practice in deep learning \cite{mamiri} to improve image segmentation; our ablation study confirmed its value. With TTA, the model’s Dice score improved, and false positive detections decreased compared to no augmentation. This is because TTA smooths out predictions and mitigates the randomness and ambiguities that arise from sparse and weak labels. In fact, the augmentation helped compensate for the limited precision of the coarse annotations, improving the output's spatial consistency. Also, we employed multi-task learning to simultaneously optimize aneurysm classification and segmentation, effectively reducing false-positive predictions. In our design, the encoder’s shared feature maps feed both a pixel‑wise segmentation head and a patch‑level classification head, enforcing that features beneficial for one task regularize the other. Our ablation study (Tables~\ref{tab:detect_ablation} and ~\ref{tab:seg_ablation}) showed that removing this interaction increases false positive rates and degrades segmentation results, underscoring the synergy of joint optimization.

However, reductions in the false positive rate come at the expense of sensitivity. Architecturally, our vesselness priors sharpen the network's attention to well-defined vasculature, which helps precision but potentially masks small or low-contrast aneurysms. Likewise, test time augmentation smooths predictions to suppress spurious detections, yet it can also eliminate low-confidence true positives. To better balance this trade-off, we aim to explore uncertainty-based loss functions in future works, to optimize both sensitivity and specificity for high stakes clinical settings where the costs of missed aneurysms and false alarms are severe.

To thoroughly evaluate the detection and segmentation performance of our VP UNet, we benchmarked it against the 3D adaptation of MT-UNet by Zhu et al.~\cite{hzhu}, as well as several established baseline architectures, including the 3D U-Net, Swin UNETR, and 3D ResUnet. All models were trained under identical conditions, allowing for a fair comparison between UNet-based architectures. These experiments provided a robust context for interpreting our results and confirmed the effectiveness of the proposed enhancements. However, we trained exclusively on the Lausanne dataset, without cross‑site or cross‑modality data, which may limit generalizability to other scanners or protocols. Finding more weakly annotated aneurysm datasets could further improve the performance of our model.

It is to be noted that while our model achieved strong results on the internal dataset, all models reported a performance drop when evaluated on the external dataset. This reduction can be partially attributed to the differences in obtained MRA scans, with the ADAM dataset showing different rotation and cropping of the brain scans compared to the Lausanne dataset. Because CNNs are not inherently rotation‑invariant, differences in scan orientation can degrade performance. To mitigate this, we applied aggressive geometric augmentations during training, including random rotations and scaling, to encourage invariance to such spatial variability. To improve domain adaptation on external sets, we will explore full‑volume augmentations (rather than patch‑based) and architectural changes that better capture global context, reducing reliance on preprocessing. A recent study by Vach et al~\cite{mvach} also evaluated the reproducibility of a CNN-based aneurysm detection and segmentation model across heterogeneous datasets and reported a similar ~10\% drop in sensitivity when applying similar pre-processing steps as our model. Although they were able to improve the gap by individually cropping each image, in future work we aim to focus on improving the robustness of the VP UNet through the framework itself, to reduce the pre-processing workload.

\section{Conclusions}
\label{sec:conclusion}

In conclusion, we have presented the VP UNet, a novel 3D multi-task segmentation and detection framework for unruptured intracranial aneurysms in TOF-MRA, trained using weak supervision. By incorporating Frangi vesselness maps as soft anatomical priors, our model effectively focuses learning on vascular regions while remaining robust to vessel filter imperfections. Through the integration of multi-task learning and test-time augmentation, VP UNet achieved strong segmentation and detection performance, outperforming several established U-Net baselines despite relying only on coarse spherical labels. Evaluated on both internal and external datasets, our results demonstrate the feasibility and scalability of weakly supervised aneurysm analysis. 

\vspace{0.5cm}
\noindent \textbf{Acknowledgment}

\noindent
We acknowledge the support of the Natural Sciences and Engineering Research Council of Canada (NSERC) and the Fonds de recherche du Québec–Nature et technologies (\href{https://doi.org/10.69777/296459}{https://doi.org/10.69777/296459} and \href{https://doi.org/10.69777/361263}{https://doi.org/10.69777/361263}). Y.X. is supported by the Fond de la Recherche du Québec – Santé (FRQS-chercheur boursier Junior 1) and Parkinson Quebec (\href{https://doi.org/10.69777/330745}{https://doi.org/10.69777/330745}).

{
    \small
    \bibliographystyle{ieeenat_fullname}
    \bibliography{main}
}

\end{document}